\documentclass[11pt]{article}

\usepackage[preprint]{acl}

\usepackage{times}
\usepackage{latexsym}
\usepackage[T1]{fontenc}
\usepackage[utf8]{inputenc}
\usepackage{microtype}
\usepackage{inconsolata}

\usepackage{amsmath, amssymb, amsthm}
\usepackage{booktabs}
\usepackage{graphicx}
\usepackage{xcolor}
\usepackage{hyperref}
\usepackage{url}

\usepackage{CJKutf8}
\newcommand{\cn}[1]{\begin{CJK}{UTF8}{gbsn}#1\end{CJK}}

\newcommand{\PP}{\mathbb{P}}
\newcommand{\one}{\mathbf{1}}

\newcommand{\ndcgten}{\mathrm{NDCG}@10}
\newcommand{\ndcgfive}{\mathrm{NDCG}@5}
\newcommand{\ndcgtwenty}{\mathrm{NDCG}@20}

\newcommand{\chargeq}{c_q}
\newcommand{\chargecand}{c_d}
\newcommand{\charge}{c}
\newcommand{\rel}{r}

\newcommand{\drop}{\Delta_{\text{drop}}}

\newcommand{\holm}{p_{\text{Holm}}}

\title{Charge as a Construct-Validity Factor in Chinese Legal Case Retrieval: \\
       A Cross-Benchmark Audit}

\author{
  Yao Liu\textsuperscript{1,2} \quad Tien-Ping Tan\textsuperscript{2} \quad Zhilan Liu\textsuperscript{3} \\[3pt]
  \textsuperscript{1}The Engineering and Technology College, Chengdu University of Technology, Leshan, China \\
  \textsuperscript{2}School of Computer Sciences, Universiti Sains Malaysia, Penang, Malaysia \\
  \textsuperscript{3}Department of Art and Design, The Engineering and Technology College, \\
  Chengdu University of Technology, Leshan, China \\[2pt]
  Correspondence: \texttt{tienping@usm.my}
}

\begin{document}
\maketitle

\begin{abstract}
Chinese Legal Case Retrieval (LCR) benchmarks grade a reference judgment relevant to a query fact pattern when their legal characterizations align, and strong neural systems now reach $\ndcgten$ of $0.85$--$0.88$. We show that most of the gap between BM25 and the best trained system is recoverable with no retrieval model at all: ranking candidates only by whether they share the query's primary charge, broken by BM25, closes $99.2\%$ of it on LeCaRDv2, with no detectable difference from the best-trained system itself. This is less a hidden flaw we uncover than a property of the benchmark's design that we quantify: LeCaRDv2 defines top relevance through a match in the crime's \emph{key constitutive elements}, which largely encode the charge, so same-charge cases are relevant almost by construction (relevance lift $4.49$; charge-to-relevance macro-AUC $0.871$). Holding charge fixed isolates the non-definitional remainder: a small within-charge residual ($+0.026$ $\ndcgten$, charge-cluster-bootstrap CI excluding zero, about a quarter of the trained reranker's advantage over BM25) is what survives, so the headline score is mostly but not entirely charge by construction. The effect, however, is not uniform, and that contrast is the substance of our finding. The same charge-only rule recovers just $84.3\%$ of the gap on LeCaRDv1 and falls \emph{out of spec} on CAIL2022, trailing the strongest trained reranker, with the charge-to-relevance signal weakening in step (macro-AUC $0.871/0.759/0.728$); a predicted-charge cascade reproduces $76.6\%$ of the LeCaRDv2 gap but does not transfer. The same construct is cashable one stage earlier: in an exploratory first-stage positive control, injecting a zero-training charge-pool channel lifts LeCaRDv2 recall ($\mathrm{R}@100$ $+0.025$, wrong-charge controls hurt), showing that $\mathrm{R}@100$ can be increased by exploiting charge membership; we report this as evidence for the confound, not as better retrieval, a retrieval method, or a novelty claim. Charge is thus a high-leverage construct-validity factor \emph{at the benchmark level}: not a uniform explanation of $\ndcgten$, and not evidence that any system relies on charge. We instantiate established construct-validity and partial-input checks as a reusable, charge-controlled evaluation protocol (CCE) for Chinese LCR, rather than proposing a new audit method; on all three benchmarks its own triggers produce no robust confirmatory result (on CAIL2022, a single depth-specific occlusion signal we treat as exploratory), behaving as designed. We release the scripts, schema, and protocol so future benchmarks can be screened for this factor before their $\ndcgten$ is read as legal-reasoning ability.
\end{abstract}

\section{Introduction}
\label{sec:intro}

A strong $\ndcgten$ on Chinese Legal Case Retrieval (LCR) is naturally read as legal-reasoning ability. The leading benchmarks (LeCaRDv2~\citep{li2023lecardv2}, LeCaRDv1~\citep{ma2021lecard}, and CAIL2022 stage-2~\citep{cail2022}) rank reference judgments for a given fact pattern and score systems with $\ndcgten$ over graded relevance, and recent strong systems reach $\ndcgten$ up to $\approx 0.85$--$0.88$: KELLER~\citep{deng2024keller} on the SAILER backbone, Qwen3-8B-Reranker~\citep{qwen3embedding}, BGE-M3~\citep{chen2024bgem3}, and SAILER itself~\citep{li2023sailer} (GLIER~\citep{glier2026} reports a higher MAP but not $\ndcgten$). Yet it is unclear how much of that number reflects within-charge legal-reasoning capacity rather than across-charge separation that any charge classifier could provide, a distinction a single leaderboard cannot draw.

Existing LCR evaluation does not separate two interpretations of strong $\ndcgten$. \textbf{Charge-as-anchor:} a fact pattern about an intentional-injury crime should retrieve intentional-injury cases, so leveraging charge identity is a legitimate retrieval signal. \textbf{Charge-as-label-leak:} if relevance grades collapse to same-versus-different charge by pool construction, then $\ndcgten$ rewards charge classification more than legal reasoning. The two are observationally indistinguishable on a single leaderboard. The question is not whether charge information \emph{should} be available to a retrieval system (it manifestly helps real downstream tasks), but whether $\ndcgten$ on existing benchmarks measures \emph{predominantly} charge matching.

Across LeCaRDv2, LeCaRDv1, and CAIL2022 stage-2, charge identity is a high-leverage \textbf{benchmark-level construct-validity factor}, but not a uniform explanation of $\ndcgten$. The standard score is \emph{largely explainable by charge} on LeCaRDv2 and only partly so on the other two benchmarks; we establish this with four \emph{model-free} or training-light measurements per benchmark (Figure~\ref{fig:hero-dashboard}). To make the test reusable, we release \textbf{CCE}, a charge-controlled evaluation packet that instantiates established construct-validity checks~\citep{gururangan2018,poliak2018,shao2026} for the charge-as-anchor versus charge-as-label-leak question, with documented decision rules. A future Chinese LCR benchmark can then be screened for this factor before its $\ndcgten$ is read as legal-reasoning ability. \textbf{We do not claim charge-specific reliance by KELLER or by model families}; our claim is benchmark-level, not system-level.

\paragraph{Contributions.}
\begin{enumerate}\setlength{\itemsep}{0pt}\setlength{\parskip}{0pt}
\item \textbf{Cross-benchmark audit.} A model-free charge-overlap oracle recovers $99.2\%$ of the BM25-to-KELLER $\ndcgten$ gap on LeCaRDv2 (closure CI $[86.9, 114.3]\%$, not detectably different from best-trained, a non-rejection rather than an equivalence test) and $84.3\%$ on LeCaRDv1, but is \textsc{out of spec} on CAIL2022: an anchor-dependent, $n{=}40$ verdict in which the oracle falls below \emph{only} Qwen3-8B-Reranker (\S\ref{sec:results-sufficiency}).
\item \textbf{Non-definitional within-charge residual.} Because the closure above is largely charge-by-construction, we isolate the part of the trained advantage that is not: holding charge fixed (same-charge candidate pool), KELLER's $+0.103$ $\ndcgten$ advantage over BM25 collapses to a within-charge residual of $+0.026$ (charge-cluster-bootstrap CI $[+0.008, +0.043]$, excludes zero; $+0.038$ on the grade-contrast subset). About a quarter of the advantage is genuine within-charge discrimination; the rest is charge-level ordering. This is the audit's one positive that does not follow from the relevance rubric (\S\ref{sec:results-sufficiency}).
\item \textbf{Label-recoverability diagnostic.} A high same-charge relevance probability is partly \emph{definitional} under LeCaRDv2's Characterization rubric (\S\ref{sec:related}); the informative quantity is the macro-AUC of charge $\to$ relevance, $0.871$ / $0.759$ / $0.728$, whose cross-benchmark gradient (point estimates; the v1 and CAIL2022 AUC CIs overlap) is not implied by any single dataset's rubric (same-versus-different-charge lift $4.49$ / $2.00$ / $1.76$; LeCaRDv2 macro-AUC covers $112/160$ pools, since all-positive pools, the most charge-homogeneous, are undefined for AUC and excluded, so $0.871$ is a conservative estimate; \S\ref{sec:results-construction}).
\item \textbf{Reproducibility diagnostic.} A predicted-charge plus BM25 cascade, with no retriever training, closes $76.6\%$ / $6.6\%$ / $-25.5\%$ of the BM25-to-best-trained gap (\S\ref{sec:results-cascade}).
\item \textbf{First-stage positive control.} On an exploratory $159$-query first-stage split, the charge construct is cashable one stage before reranking: a zero-training channel fusing BM25, a dense retriever, and a self-predicted-charge-restricted BM25 list lifts first-stage $\mathrm{R}@100$ on LeCaRDv2 by $+0.0250$ (Bonferroni $m{=}6$ CI $[+0.0077, +0.0412]$ after post-hoc selection over six variants) while wrong-charge controls strictly hurt. Dense-rank provenance is unverified; both arms share the same fixed dense list, so the paired delta is not a dense-quality artifact, though the self-predicted charge that builds the charge-pool list does itself draw on a BM25-plus-dense top-20 vote. We report this as evidence that injecting charge exploits the first-stage confound, not as better retrieval, a retrieval method, or a novelty claim (\S\ref{sec:results-carrf}).
\item \textbf{Released evaluation packet (CCE).}\footnote{Code, locked protocol, and per-query result data: \url{https://github.com/usmliuyao/cce-chinese-lcr-audit}.} A \emph{reusable, documented} charge-controlled evaluation packet for Chinese LCR: charge-stratified $\ndcgten$, charge-name occlusion, and charge-clustered significance testing, with documented decision rules (\S\ref{sec:method-cce}, Appendix~\ref{app:cce-protocol}: primary $k{=}10$, trigger thresholds, multi-charge handling, out-of-spec rule, reporting template). It \emph{instantiates established construct-validity checks}~\citep{shao2026,freiesleben2025} (metadata-only screening, evidence-occlusion intervention, charge-stratified reporting) specialized to legal case ranking; we claim the packaging and the legal-IR specialization, not a new audit method. On the three benchmarks here, its own triggers produce no robust confirmatory result: a small-strata, contamination-sensitive top-3 reweighting alert on LeCaRDv2 (not a significance flip, \S\ref{sec:results-strat}), and a single depth-specific CAIL2022 counterfactual signal we treat as exploratory rather than confirmed (\S\ref{sec:results-cf}). The packet behaves as designed.
\end{enumerate}

\begin{table*}[t]
\centering
\small
\begin{tabular}{@{}p{0.28\textwidth} p{0.42\textwidth} p{0.22\textwidth}@{}}
\toprule
\textbf{What we claim} & \textbf{Evidence (this paper)} & \textbf{What we do \emph{not} claim} \\
\midrule
Same-charge primary-key ranking is a strong \emph{sufficiency diagnostic} on LeCaRDv2. &
A model-free charge-overlap oracle closes $99.2\%$ of the BM25-to-best-trained $\ndcgten$ gap; the closure CI $[86.9,114.3]\%$ includes $100\%$ and the oracle$-$best gap CI contains zero (\S\ref{sec:results-sufficiency}). &
A new audit theory or method: the audit logic is established~\citep{gururangan2018,poliak2018,shao2026}. \\
\midrule
The charge effect is \emph{benchmark-heterogeneous}, not universal. &
Closure $99.2\%$ / $84.3\%$ on LeCaRDv2 / v1 but \textsc{out of spec} on CAIL2022 (\S\ref{sec:results-sufficiency}). &
That Chinese LCR benchmarks are globally invalid. \\
\midrule
Relevance labels are \emph{charge-predictable}, partly by the v2 rubric's design. &
Same-vs-different-charge lift $4.49$ / $2.00$ / $1.76$; macro-AUC $0.871$ / $0.759$ / $0.728$, a non-trivial cross-benchmark gradient (\S\ref{sec:results-construction}). &
That this is a hidden artifact we uncovered, or that any system \emph{merely matches charge} (our claim is benchmark-level). \\
\midrule
A small \emph{within-charge signal} survives charge control (the non-definitional residual). &
Holding charge fixed, KELLER's $+0.103$ full-pool $\ndcgten$ advantage over BM25 shrinks to $+0.026$ (charge-cluster-bootstrap CI $[+0.008,+0.043]$, excludes zero); $+0.038$ grade-contrast (\S\ref{sec:results-sufficiency}). &
That the residual is large or reflects a charge mechanism; ${\sim}3/4$ of the advantage is charge-level ordering. \\
\midrule
We release reusable \emph{charge-controlled evaluation infrastructure} for Chinese LCR. &
Scripts, score/qrels schema, charge-stratified $\ndcgten$, charge-name occlusion, and charge-cluster bootstrap (\S\ref{sec:method-release}). &
A deployable retrieval system: the oracle is a diagnostic baseline, not a retriever. \\
\bottomrule
\end{tabular}
\caption{\textbf{Claims, evidence, and explicit non-claims.} The paper's contribution is empirical and infrastructural: a model-free charge-overlap diagnostic and a released charge-controlled evaluation packet for Chinese LCR. We do not claim novelty of the underlying construct-validity audit logic.}
\label{tab:claims}
\end{table*}

Table~\ref{tab:claims} states these claims alongside the evidence for each and our explicit non-claims: the contribution is empirical and infrastructural (a model-free charge-overlap diagnostic and a released charge-controlled evaluation packet for Chinese LCR), not a new audit method.

The rest of the paper is organized as follows. Section~\ref{sec:related} positions CCE against charge-aware Chinese LCR systems, construct-validity work in NLP, and counterfactual evaluation. Section~\ref{sec:motivation} formalizes \emph{correction versus over-control}. Sections~\ref{sec:method} and~\ref{sec:results} describe the audit framework and report cross-benchmark results. Section~\ref{sec:conclusion} concludes; a Limitations section covers the three-benchmark scope and (partly confirmed) baseline contamination.

\section{Related Work}
\label{sec:related}

\paragraph{Charge-aware Chinese LCR systems.}
Recent Chinese LCR systems treat charge as a useful juridical signal. KELLER~\citep{deng2024keller} uses an LLM-guided slot decomposition over the SAILER~\citep{li2023sailer} backbone, with charge labels appearing as slot keys in pre-cached crime-fact dictionaries. GLIER~\citep{glier2026}, a recent strong system reporting higher MAP, fuses charge as a feature with lexical signals and reports a $-16.39$ MAP drop when the charge feature is ablated, establishing charge as a model-side ``gatekeeper'', a \emph{model-feature} estimand distinct from our \emph{model-free benchmark label-recoverability} measurement. CaseLink~\citep{caselink2024} builds a charge-graph contrastive objective; LeDSGra~\citep{ledsgra2024} layers Chinese-statute hierarchy over case graphs. \emph{All four incorporate charge as a model-side signal during training or representation; they do not measure whether benchmark labels themselves are recoverable from charge-primary ranking with no learned reranker.} Our work is orthogonal: we treat charge as a \emph{benchmark-level} construct-validity factor and ask when its use is correction versus over-control.

\paragraph{Benchmark validity and dataset artifacts in NLP.}
Annotation artifacts repeatedly undermine leaderboard interpretation. Partial-input and hypothesis-only baselines~\citep{gururangan2018,poliak2018,hans2019,nivenkao2019,kaushik2018much} show that benchmark labels are often recoverable from a label-correlated subset of the input without the target competence. Underpowered comparisons can themselves manufacture apparent leaderboard gains~\citep{card2020power}, which motivates the charge-cluster-bootstrap confidence intervals and explicit power caveats we report throughout. \citet{bowmandahl2021}, \citet{rajipress2021}, and \citet{freiesleben2025} formalize benchmark construct validity (with \citet{bean2025measuring} documenting validity breaches across hundreds of recent benchmarks), and \citet{shao2026} packages metadata-only screening, an evidence-intervention statistic, and reader-strength calibration into a general weak-label audit. \citet{shao2026} also cautions that metadata predictability alone does not establish evidence dependence; consistent with that caution we do not infer system-level evidence-independence from charge predictability, and our charge-name occlusion (\S\ref{sec:results-cf}) and within-charge residual (\S\ref{sec:results-sufficiency}) are instances of the evidence-intervention and non-definitional-remainder checks his protocol recommends, showing the score is mostly but not entirely charge by construction. Construct-validity audits of \emph{retrieval} benchmarks' relevance labels, with corrected protocols, also exist~\citep{fire2023}, and behavioral probes characterize what neural rankers actually rely on~\citep{macavaney2022abnirml}. We contribute a legal-IR instance of this lineage: a model-free charge-overlap baseline on graded \emph{ranking} ($\ndcgten$) rather than classification accuracy, asking whether Chinese LCR \emph{labels} reward charge identity above legal-reasoning signal. We do not claim the audit machinery is new; our delta is the ranking-metric instantiation, the cross-benchmark heterogeneity, and the legal-domain specialization.

\paragraph{Chinese LCR benchmark construction and qrels pooling.}
LeCaRDv1~\citep{ma2021lecard}, LeCaRDv2~\citep{li2023lecardv2}, and CAIL2022~\citep{cail2022} all construct candidate pools via automatic retrieval over public Chinese-judgment corpora before manual relevance grading. Two distinct mechanisms then tie charge to relevance, and we separate them explicitly. \emph{By annotation rubric:} LeCaRDv2's guideline~\citep{li2023lecardv2} defines its primary aspect, \emph{Characterization} relevance, through the case's ``key constitutive elements of the crime (key elements)'' together with key circumstances, and assigns the top relevance levels only when the key elements match. Because the key constitutive elements of an offense substantially encode its charge, the near-saturated same-charge relevance probability we report ($\PP(\rel{\geq}2\mid\chargeq{=}\chargecand){=}0.97$ on LeCaRDv2) is in large part \emph{definitional} under the published rubric, not a latent artifact we uncover. \emph{By pooling:} because the source documents are organized by statutory charge, lexical pooling additionally shapes the \emph{different}-charge base rate, a domain-specific instance of the pooling-bias artifacts long studied in retrieval test collections~\citep{arabzadeh2022shallow}, where pooling depth and contributing systems determine which documents are ever judged. Our contribution is therefore not to reveal that charge predicts relevance; the rubric says so by design. It is to \emph{quantify} how far that by-design structure propagates into $\ndcgten$ (the sufficiency oracle, \S\ref{sec:method-sufficiency}), and how much it varies across benchmarks that instantiate the rubric to differing degrees (the construction probe's macro-AUC gradient, \S\ref{sec:method-construction}).

\paragraph{Group-control and counterfactual evaluation.}
FairLex~\citep{chalkidis2022fairlex} \emph{evaluates} group-robust algorithms (Group DRO, IRM, Adversarial Removal) for legal-document classification across protected attributes. \emph{FairLex applies group-control to legal-classification training; we apply group-control to retrieval-side benchmark evaluation.} CCE's counterfactual analog uses text-level perturbation (charge-name occlusion) rather than the policy-level perturbations of fairness-oriented IR evaluation; we frame it as a construct-validity audit, not a fairness intervention.

\paragraph{Baseline contamination context.}
Several systems we evaluate have prior exposure to public Chinese-legal corpora. BGE-M3~\citep{chen2024bgem3} lists LeCaRDv2 explicitly among its disclosed Chinese fine-tuning datasets, a confirmed supervised contamination on this benchmark; the Qwen3 series~\citep{qwen3} does not disclose its pretraining corpus; SAILER~\citep{li2023sailer} pretrains on Chinese criminal-law text. No main system is verified clean of LeCaRDv2 exposure. This strengthens rather than weakens our motivation for benchmark-level evaluation tools that do not assume clean baselines, and is why our load-bearing diagnostics do not rely on the contaminated system: the oracle ranking rule and the construction probe are model-free, and oracle closure anchors on KELLER/Qwen3, not BGE-M3 (\S\ref{sec:method-sufficiency},~\S\ref{sec:method-construction}).

\section{Motivation: Charge as Correction vs.\ Over-Control}
\label{sec:motivation}

Prior work uses charge as a model-side signal; we ask whether $\ndcgten$ on existing Chinese LCR benchmarks measures \emph{predominantly} that signal. Charge labels are routinely available for adjudicated Chinese criminal judgments and have been widely used as supervisory signal~\citep{deng2024keller,glier2026}, graph structure~\citep{caselink2024}, and ontological hierarchy~\citep{ledsgra2024}. The question is therefore not whether charge information \emph{should} be available to a retrieval system (it manifestly helps real downstream tasks) but whether the metric leaves room for the within-charge legal reasoning that distinguishes a strong retriever from a charge-only ranking rule. This anchor-versus-reasoning split is the benchmark's own: LeCaRDv2's Characterization rubric~\citep{li2023lecardv2} factorizes relevance into matching the crime's \emph{key constitutive elements} (which encode charge) and matching \emph{key circumstances} (the within-charge facts), so the question we pose is which of the rubric's two factors $\ndcgten$ ends up rewarding.

Our cross-benchmark evidence (Section~\ref{sec:results}) shows the answer depends on the benchmark. On LeCaRDv2, $\PP(\rel \geq 2 \mid \chargeq{=}\chargecand)=0.97$ versus $\PP(\rel \geq 2 \mid \chargeq{\neq}\chargecand)=0.22$ (lift $4.49$), and a model-free charge-overlap oracle recovers $99.2\%$ of the BM25-to-KELLER $\ndcgten$ gap. On CAIL2022 stage-2 the same oracle falls below Qwen3-8B-Reranker's $\ndcgten$, indicating substantial retrieval-specific headroom beyond the charge-primary oracle.

We therefore frame CCE as a \textbf{correction-versus-over-control test}, not an anti-charge protocol. Charge-stratified evaluation quantifies how much of a system's headline $\ndcgten$ comes from cross-charge ranking (where a charge-only rule explains much of the separation) versus within-charge ranking (where legal-reasoning signal lives). Charge control is \textbf{correction} when most of the standard gap collapses under stratification (LeCaRDv2: gold-charge oracle $\approx$ best-trained) and \textbf{over-control} when differentiation beyond the charge-primary oracle already drives the score (CAIL2022: oracle below best-trained). The counterfactual analog re-ranks under charge-name occlusion, testing whether rankings change when explicit charge cues are removed, without attributing the effect to any model mechanism. Section~\ref{sec:method} operationalizes this through four measurements and the CCE protocol with explicit decision rules.

\section{Method}
\label{sec:method}

We define four cross-benchmark measurements (\S\ref{sec:method-sufficiency}--\S\ref{sec:method-cascade}) and the Charge-Controlled Evaluation (CCE) protocol (\S\ref{sec:method-cce}--\S\ref{sec:method-release}). All measurements are computed on LeCaRDv2 test, LeCaRDv1 test, and CAIL2022 stage-2 (40-query hard pool), using the released qrels with KELLER's $\ndcgten$ formula~\citep{deng2024keller} ($\mathrm{gain}(\rel){=}2^{\rel-1}$ for $\rel{\geq}1$, $0$ otherwise; did-desc tie-break) to maintain comparability with published numbers.

\textbf{Evaluated panel.} Our reproducible panel has six systems: the \emph{five-system main family} (BM25, BGE-M3, SAILER, chinese-roberta-wwm-ext, Qwen3-8B-Reranker) over which the charge-controlled evaluation's $\alpha{=}0.05$ Holm-corrected FWER is computed, plus KELLER as an external diagnostic (\S\ref{sec:method-cce}). GLIER~\citep{glier2026} reports stronger LeCaRDv2 numbers but uses a non-comparable pipeline; we therefore use \emph{best-trained-in-panel} to mean the per-benchmark argmax over the full six-system panel \emph{including} KELLER as a published reference point. This anchor is data-determined: it is KELLER on LeCaRDv2/v1 and Qwen3-8B-Reranker on CAIL2022, because KELLER scores below Qwen3-8B there ($0.802$ vs.\ $0.851$), not because it is excluded. KELLER is excluded only from the five-system Holm FWER \emph{significance} family (\S\ref{sec:method-cce}), a separate role. The cross-benchmark heterogeneity is invariant to this convention: excluding KELLER entirely, the oracle still \emph{over}-recovers on LeCaRDv2 (oracle $0.876$ vs.\ next-best Qwen3-8B $0.817$) and CAIL2022 stays out of spec.

\textbf{Per-protocol denominators.} LeCaRDv2 ($n{=}160$) and CAIL2022 ($n{=}40$) agree across protocols; LeCaRDv1 uses $n{=}104$ (oracle and cascade, after classifier-label drops), $107$ (construction probe, no classifier dependency), and $103$ (charge-controlled evaluation, after pool-consistency), per Appendix~Table~\ref{tab:denominators}.

\subsection{Sufficiency oracle}
\label{sec:method-sufficiency}

\textbf{Definition.} For each query $q$, rank candidates $d$ by $\one[\chargecand{=}\chargeq]$ as the primary key (no trained model, no learned representation), with BM25~\citep{robertson1995bm25} as the deterministic tie-break score within each same-charge partition. Compute $\ndcgten$ over the released qrels; the per-query candidate pool includes graded non-relevant candidates, so BM25 retains full dynamic range (its $\ndcgten$ reaches $0$ on some queries) and the pool is not positives-saturated. We refer to this as the \emph{model-free charge-overlap oracle}; ``model-free'' denotes the absence of any learned reranker, while BM25 enters only as a deterministic lexical tie-break inside same-charge strata.

\textbf{Claim.} The oracle uses only gold primary-charge labels as the ranking primary key; no learned reranking signal enters the partition order. BM25 affects ordering only within each same-charge partition, never across charges. On benchmarks where the oracle's $\ndcgten$ approaches the best-trained-in-panel system, we interpret the oracle as a \emph{charge-primary BM25-tie-break diagnostic baseline} (not an absolute upper bound; a learned within-charge ranker could in principle exceed it) and conclude that the standard score is largely explainable by charge identity plus a deterministic lexical tie-break.

\subsection{Construction probe}
\label{sec:method-construction}

The construction probe quantifies how charge-predictable the relevance labels are by pool construction, using two label-only statistics. For each benchmark, we estimate $\PP(\rel{\geq}2 \mid \chargeq{=}\chargecand)$ and $\PP(\rel{\geq}2 \mid \chargeq{\neq}\chargecand)$ over the released qrels. The \textbf{lift} is the ratio of the first to the second; \textbf{macro-AUC} is the area under the binary ROC curve $\one[\chargeq{=}\chargecand] \to \one[\rel{\geq}2]$ macro-averaged over queries. Macro-AUC is undefined for all-positive pools (no relevance contrast); on LeCaRDv2 this leaves coverage $112/160$ ($47$ all-positive plus $1$ all-negative pool), which formally fails a pre-registered $0.9$ coverage gate ($144/160$). We report macro-AUC regardless, with full disclosure: the excluded pools are the most charge-homogeneous, so $0.871$ is a \emph{conservative estimate} (an argument from the excluded pools' homogeneity, not a formal bound); the $47/160$ all-positive top-heaviness is itself a construction finding; and a coverage-free pooled-AUC ($0.924$) corroborates the same direction.

\subsection{Reproducibility cascade}
\label{sec:method-cascade}

\textbf{Classifier.} We train a multi-label charge classifier on chinese-roberta-wwm-ext~\citep{cui2020roberta} using LeCaRDv2's 640 train queries with their gold primary-charge labels (89 unique charges). Per-class \texttt{pos\_weight} rebalancing addresses class imbalance; eight epochs at $\mathrm{lr}{=}2{\times}10^{-5}$. For LeCaRDv1 and CAIL2022 we apply the same classifier with no fine-tuning; predicted labels outside the 89-charge LeCaRDv2-train vocabulary fall back to the closest matching label by surface form, and queries with no candidate match are excluded.

\textbf{Cascade.} At inference, rank candidates by $\one[\widehat{\charge}_d{=}\widehat{\charge}_q]$ as the primary key, with BM25 as the tie-break, where $\widehat{\charge}$ is the classifier's top-1 prediction. The cascade requires \emph{no retriever training}; only the small charge classifier is learned.

\textbf{Scope.} The cascade is a stress test of \emph{portable charge prediction plus BM25}: when the classifier transfers well (LeCaRDv2 in-distribution), it operationalizes the oracle result with predicted instead of gold labels; when transfer fails (LeCaRDv1, CAIL2022), the cascade conflates two effects (benchmark label-recoverability and classifier domain shift), and its closure should be read as the joint outcome, not as a clean benchmark-only diagnostic.

\textbf{Fence.} The $99.2\%$ recovery quoted in \S\ref{sec:results-sufficiency} applies only to the gold-charge oracle (continuous over true labels); it does \emph{not} apply to the cascade, which uses predicted labels plus a discrete tie-break.

\subsection{First-stage positive control}
\label{sec:method-carrf}

\textbf{Construction.} A zero-training probe of whether the charge confound is cashable at first-stage recall, not only at rerank. We fuse three ranked lists with weighted reciprocal-rank fusion ($k{=}60$): (i) BM25 over fact text, (ii) a dense retriever (SAILER), and (iii) a charge-pool list. The charge pool is \emph{self-predicted} with no labels at inference: the two most-voted charges among the fused top-$20$ of (i)$+$(ii) (rank-weighted vote over corpus charge metadata) define a charge set, and list (iii) is the global BM25 ranking restricted to documents carrying those charges. No component is trained for this task.

\textbf{Controls and attribution.} Two placebo arms replace the self-predicted charges with the next-ranked (offset) charges or random size-matched charges; an exact Shapley decomposition over the three fusion channels attributes the gain per query. Evaluation is macro $\mathrm{R}@100$ over the $159$ LeCaRDv2 test queries with available dense ranks; CIs are paired query bootstrap ($B{=}10000$).

\textbf{Scope.} This is a \emph{positive control}, not a retrieval method: it tests whether injecting self-predicted charge membership cashes first-stage $\mathrm{R}@100$ (it does; wrong-charge controls hurt), which evidences benchmark sensitivity to the charge construct rather than better legal retrieval. We make no novelty or retrieval-advance claim; the construction recombines reciprocal-rank fusion, pseudo-relevance feedback, and category-restricted retrieval, and is reported only as stage-independent corroboration of the confound. The single split is exploratory (the self-predicted-charge variant was chosen post-hoc over six tried; Bonferroni $m{=}6$ CIs reported in \S\ref{sec:results-carrf}). Dense-rank provenance is unverified; this affects absolute recall, not the paired delta, because both fusion arms share the dense channel.

\subsection{CCE protocol}
\label{sec:method-cce}

\textbf{Definition (3 sentences).} \textbf{CCE} evaluates case retrieval under explicit charge control by reporting (i) \emph{charge-stratified $\ndcgten$} (macro-averaged across charge strata, $N^{\mathrm{F}}$) and (ii) \emph{counterfactual occlusion drop $\drop$} (baseline $N^{\mathrm{F}}$ minus occluded $N^{\mathrm{F}}$) per system. CCE uses a \emph{cluster bootstrap over queries grouped by charge}~\citep{cameronmiller2015cluster} to produce uncertainty and significance tests valid under charge clustering. The counterfactual analog re-scores systems after replacing charge-name spans in both the query and candidate fact fields with a fixed placeholder, to measure the sensitivity of $\ndcgten$ to explicit charge-name surface forms.

\textbf{Formal definitions.} Let $\mathcal{S}$ be charge strata, $\mathcal{Q}_s$ the queries in stratum $s$, and $N_s(\pi){=}|\mathcal{Q}_s|^{-1}\sum_{q\in\mathcal{Q}_s} \ndcgten(q,\pi)$ the per-stratum mean. The charge-stratified NDCG@10 is $N^{\mathrm{F}}(\pi){=}|\mathcal{S}|^{-1}\sum_{s} N_s(\pi)$; counterfactual drop is $\drop(\pi){=}N^{\mathrm{F}}(\pi){-}N^{\mathrm{F}}(\pi^{\text{occ}})$. A charge-clustered bootstrap ($B{=}10000$, seed $20260528$; the released pre-registration lists $B{=}2000$ as the planned resample count for the future held-out evaluation, so the present audit's larger $B$ only sharpens CIs and $p$-values, leaving point estimates unchanged) produces paired $p$-values for $N^{\mathrm{F}}(\pi_A){-}N^{\mathrm{F}}(\pi_B)$ and $\drop^A{-}\drop^B$ over $C(5,2){=}10$ main pairs, Holm-corrected at $\alpha{=}0.05$.

\textbf{Non-penalization clause.} CCE does not penalize charge use \emph{per se}; it separates ``performance explained by charge-primary ranking (BM25 tie-break)'' from ``performance within charge strata.'' A system can score well if it improves ranking within the same-charge set.

\textbf{Scope of adjudication.} CCE provides a comparable diagnostic separation between charge-as-anchor and charge-as-label-leak; it does \emph{not} normatively adjudicate which regime a benchmark is in without task-design and statute-coverage context.

\textbf{Decision rules.} Table~\ref{tab:cce-rules} summarizes the five rules; full text in Appendix~\ref{app:cce-protocol}. The first four were specified before the cross-benchmark results were computed; the fifth rule's tri-state thresholds were fixed during analysis (Appendix~\ref{app:cce-protocol}) and are pre-registered for future-benchmark use, descriptive (not confirmatory) on the three benchmarks here.

\begin{table}[t]
\centering
\small
\begin{tabular}{@{}p{0.94\linewidth}@{}}
\toprule
\textbf{(1) Primary depth.} $k{=}10$ primary; $k{\in}\{5,20\}$ as sensitivity. \\
\midrule
\textbf{(2) Stratified trigger.} A Holm-corrected~\citep{holm1979} significance flip in the C$(5,2){=}10$-pair main FWER family at $\alpha{=}0.05$, OR a top-3 rank reversal between standard and charge-stratified $\ndcgten$. \\
\midrule
\textbf{(3) Occlusion trigger.} A Holm-corrected significant differential drop in the same 10-pair main FWER family under charge-name occlusion; KELLER excluded. \\
\midrule
\textbf{(4) Multi-charge handling.} Primary stratum = first gold charge per query; all-charges fractional-membership as sensitivity. \\
\midrule
\textbf{(5) Out-of-spec + reporting (tri-state).} Using gap $g{=}\ndcgten^{\mathrm{best}}{-}\ndcgten^{\mathrm{oracle}}$ and closure $\kappa$: \emph{Within-band} if $g{\leq}\varepsilon_{\mathrm{eq}}{=}0.005$; \emph{Partial} if $g{>}\varepsilon_{\mathrm{eq}}$ and $\kappa{\geq}80\%$; \textsc{out of spec} if $g{>}\varepsilon_{\mathrm{eq}}$ and $\kappa{<}80\%$, excluded from the primary sufficiency roll-up. Report the 7-table template (Appendix~\ref{app:cce-protocol}). \\
\bottomrule
\end{tabular}
\caption{\textbf{CCE decision rules.} The first four were specified before computing results; the fifth's thresholds fixed during analysis and pre-registered for future benchmarks (Appendix~\ref{app:cce-protocol}).}
\label{tab:cce-rules}
\end{table}

\textbf{Counterfactual occlusion mechanics.} The occlusion test replaces charge-name spans in both query and candidate fact fields with a fixed placeholder \cn{[罪名]}, using a hand-curated 258-term whitelist assembled from the PRC criminal-law statute index and the union of charge labels in LeCaRDv2/v1/CAIL2022 qrels. Statute numbers and procedural strings are not occluded; each system re-ranks the masked corpus under its native pipeline. The probe is sufficient-only: it isolates explicit charge-name sensitivity but does not strip implicit charge cues (statute references, characteristic fact patterns).

\textbf{KELLER as excluded-from-FWER external diagnostic.} KELLER's pre-cached LLM-decomposed crime-fact dictionaries use charge as structural metadata (slot key). Text-level charge-name occlusion via our whitelist cannot fully strip charge from KELLER's input without re-running its original LLM-guided slot-decomposition pipeline; KELLER's counterfactual is a \emph{pipeline-internal} mask of slot prefix and body, not directly comparable to the text-level QD-occlusion used for the 5 main systems. We therefore report KELLER under the charge-controlled evaluation in a separate row as an external diagnostic and exclude it from the main FWER family.

\subsection{Released evaluation infrastructure}
\label{sec:method-release}

We release CCE as reusable evaluation \emph{infrastructure}, not a turnkey solution: scripts (\texttt{eval\_cce\_main.py}, \texttt{eval\_cce\_counterfactual.py}), the expected score and qrels JSON schema, the locked protocol document, and sample reports on 6 systems $\times$ 3 benchmarks. Future Chinese LCR benchmark authors produce their own per-system score JSONs matching the schema; the eval scripts then run the protocol unchanged.

\section{Results}
\label{sec:results}

\begin{figure*}[t]
    \centering
    \includegraphics[width=0.62\textwidth]{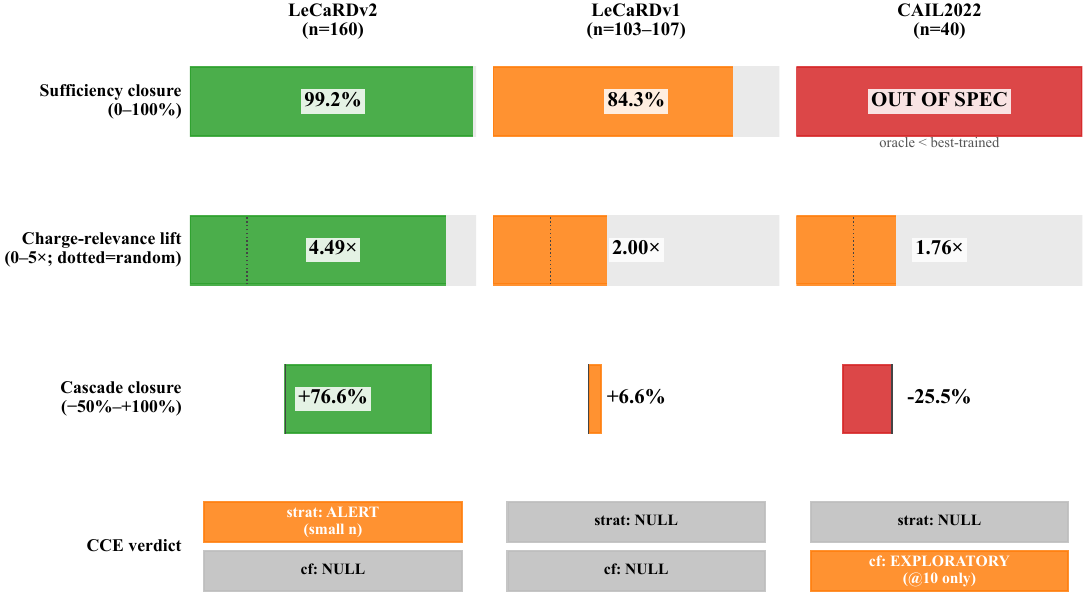}
    \caption{\textbf{Cross-benchmark heterogeneity of charge as a construct-validity factor in Chinese LCR.}
    Columns are benchmarks; rows are the four diagnostics. \textbf{Colors}: green = strong, orange = exploratory/descriptive, gray = null, red = out of spec/negative. Per-benchmark values and caveats in \S\ref{sec:results-sufficiency}--\S\ref{sec:results-cf}. We claim no charge-specific reliance by any system.}
    \label{fig:hero-dashboard}
\end{figure*}

We organize the audit around three questions. \emph{How much of the standard score a label-free charge rule recovers}: the sufficiency oracle (\S\ref{sec:results-sufficiency}), a predicted-charge cascade (\S\ref{sec:results-cascade}), and a first-stage positive control (\S\ref{sec:results-carrf}). \emph{What the relevance labels encode about charge, and what they do not}: the construction probe (\S\ref{sec:results-construction}) and the cross-charge complement it leaves behind (\S\ref{sec:results-crosscharge}). \emph{What survives once charge is held fixed}: the within-charge residual (\S\ref{sec:results-sufficiency}), the charge-stratified test (\S\ref{sec:results-strat}), and charge-name occlusion (\S\ref{sec:results-cf}). Figure~\ref{fig:hero-dashboard} summarizes the cross-benchmark pattern, and \S\ref{sec:results-summary} collects the per-benchmark verdicts.

\subsection{Sufficiency oracle, cross-benchmark}
\label{sec:results-sufficiency}

\begin{table*}[t]
\centering
\footnotesize
\setlength{\tabcolsep}{4pt}
\caption{\textbf{Sufficiency oracle across three Chinese LCR benchmarks.} The charge-overlap oracle ranks candidates by $\one[\chargecand{=}\chargeq]$ as primary key with BM25 as deterministic tie-break (no trained reranker; \S\ref{sec:method-sufficiency}). All NDCG values are at depth 10. ``Closure'' = (oracle\,$-$\,BM25)\,/\,(best-trained-in-panel\,$-$\,BM25). On CAIL2022 the oracle falls below Qwen3-8B-Reranker, so the benchmark is labelled \textsc{out of spec for sufficiency} per the out-of-spec rule (Appendix~\ref{app:cce-protocol}); this means the charge-primary oracle does not match best-trained-in-panel, not that charge identity is absent from the benchmark (the oracle still beats BM25 by $6.75$\,pp on CAIL2022). Closure percentages are computed from full-precision NDCG; recomputing from the rounded NDCG shown can differ by up to ${\sim}0.1$\,pp.}
\label{tab:sufficiency}
\begin{tabular}{@{}lccccc@{}}
\toprule
Benchmark & Best-trained system & BM25 & Best-trained & Oracle & Closure \% \\
\midrule
LeCaRDv2 ($n{=}160$) & KELLER & 0.7423 & 0.8774 & 0.8762 & 99.2\% \\
LeCaRDv1 ($n{=}104$) & KELLER & 0.6823 & 0.8449 & 0.8195 & 84.3\% \\
CAIL2022 ($n{=}40$) & Qwen3-8B & 0.7638 & 0.8511 & 0.8314 & \textsc{out of spec} \\
\bottomrule
\end{tabular}
\end{table*}

Table~\ref{tab:sufficiency} reports the sufficiency oracle (tolerances $\varepsilon_{\mathrm{eq}}{=}0.005$, $\varepsilon_{\mathrm{cl}}{=}80\%$; Appendix~\ref{app:cce-protocol} documents that these were fixed during analysis and are descriptive here, pre-registered only for future benchmarks). We read the three benchmarks as a \emph{continuous gradient}, not three sharp states. On LeCaRDv2 the model-free oracle ($0.8762$) is $0.0012$ below KELLER ($0.8774$), a $99.2\%$ closure of the BM25-to-best-trained gap; the query-bootstrap closure CI is $[86.9, 114.3]\%$ and the oracle$-$KELLER gap CI $[-0.019, +0.017]$ contains zero, so the oracle is \emph{not detectably different from best-trained} (not an equivalence test), which is the substance of the sufficiency claim, not a refutation of it. On LeCaRDv1 closure is $84.3\%$ (CI $[72.8, 98.6]\%$, straddling the $80\%$ cutoff, so the \emph{Partial} label is a descriptive threshold call, not a tested state; the gap CI $[-0.050, -0.002]$ does exclude zero, i.e.\ the v1 oracle is reliably below KELLER).

On CAIL2022 the oracle ($0.8314$) falls below Qwen3-8B-Reranker ($0.8511$), labelled \textsc{out of spec}; but this verdict is \emph{anchor-dependent}: the oracle exceeds SAILER ($0.8155$), KELLER ($0.8016$), and BM25 ($0.7638$) and falls below \emph{only} Qwen3-8B, so a non-Qwen3 anchor flips CAIL to over-recovery, and at $n{=}40$ the label is evidence-limited (no formal power calculation). (\textsc{out of spec} does not mean charge is absent: the oracle still beats BM25 by $+6.8$\,pp and the construction lift stays positive.) LeCaRDv2's over-recovery is itself anchor-invariant (the oracle exceeds every panel system except KELLER, e.g.\ vs.\ Qwen3-8B $0.8166$); the cross-benchmark conclusion does \emph{not} rest on CAIL2022's oracle label.

The heterogeneity is the headline result. LeCaRDv2's $\ndcgten$ is largely explainable by charge identity, in the sense that the model-free charge-primary oracle approaches best-trained-in-panel; on CAIL2022 this is not true. The $99.2\%$ recovery on LeCaRDv2 does \emph{not} imply that KELLER ``uses charge'': it implies the benchmark labels are recoverable from a charge-primary ranking rule with a deterministic lexical tie-break.

\paragraph{Within-charge residual: the non-definitional complement.} The $99.2\%$ closure is largely charge-by-construction (\S\ref{sec:related}); the within-charge residual isolates the part of the trained advantage that is \emph{not}, and is the one positive in this audit that does not follow from the relevance rubric. Restricting evaluation to each query's \emph{same-charge} candidate pool, where charge overlap is constant and cannot order candidates, the full-pool advantage of the trained reranker KELLER over BM25 ($+0.103$ $\ndcgten$; $0.878$ vs.\ $0.775$, all $160$ queries) collapses to a within-charge residual of $+0.026$ (charge-cluster-bootstrap CI $[+0.008,+0.043]$ over $48$ charge strata, $n{=}104$ eligible queries; query charge extracted from the query judgment text rather than the relevance labels; $+0.038$ on the stricter grade-contrast subset; released in \texttt{cce\_within\_charge\_v2\_results.json}). The residual is small, roughly a quarter of the full-pool advantage, but its cluster-bootstrap CI excludes zero: a real within-charge signal survives once charge can no longer order candidates, and it is the measurement that keeps the sufficiency result from being purely definitional. Consistent with the headline, most of the standard-$\ndcgten$ separation between a trained reranker and a lexical baseline still reflects charge-level ordering rather than within-charge discrimination, a property of the charge-structured labels and not evidence of a charge mechanism in any system. This non-definitional positive is itself benchmark-specific: under a comparable same-charge-restricted KELLER-over-BM25 comparison (same scoring protocol, not the v2 pipeline above) the residual is not distinguishable from zero on LeCaRDv1 ($+0.014$) or CAIL2022 ($-0.018$, both CIs contain zero), so even the part of the trained advantage that survives charge control is largest on LeCaRDv2, mirroring the sufficiency gradient rather than contradicting it.

\subsection{Construction probe, cross-benchmark}
\label{sec:results-construction}

\begin{table*}[t]
\centering
\small
\caption{\textbf{Construction probe across three Chinese LCR benchmarks.} Per-benchmark P(rel$\ge\,2$ $\mid$ same charge) vs.\ P(rel$\ge\,2$ $\mid$ different charge); lift is the ratio; macro-AUC is the binary classifier $\one[\chargeq{=}\chargecand]\to\one[\rel\ge2]$ averaged over queries. Lift is positive on all three benchmarks but the magnitude varies by $2.5\times$.}
\label{tab:construction}
\begin{tabular}{lcccc}
\toprule
Benchmark & P(rel$\ge$2$\,|\,$same charge) & P(rel$\ge$2$\,|\,$diff charge) & Lift & Macro-AUC \\
\midrule
LeCaRDv2 ($n{=}160$) & 0.968 & 0.215 & 4.49 & 0.871 \\
LeCaRDv1 ($n{=}107$) & 0.974 & 0.486 & 2.00 & 0.759 \\
CAIL2022 ($n{=}40$) & 0.982 & 0.558 & 1.76 & 0.728 \\
\bottomrule
\end{tabular}
\end{table*}

Table~\ref{tab:construction} reports same-vs-different-charge relevance probabilities. The near-saturated same-charge numerator ($\PP(\rel{\geq}2 \mid \chargeq{=}\chargecand){=}0.97$ / $0.974$ / $0.982$) is expected rather than surprising: it is in large part \emph{definitional} under LeCaRDv2's Characterization rubric (\S\ref{sec:related}), whose top relevance levels require the crime's key constitutive elements (which substantially encode charge) to match. The informative quantity is therefore not the lift ($4.49$ / $2.00$ / $1.76$, whose $2.5\times$ variation is carried by the different-charge base rate, a pool-composition effect) but the base-rate-robust \textbf{macro-AUC} of $\one[\chargeq{=}\chargecand] \to \one[\rel{\geq}2]$: $0.871$ / $0.759$ / $0.728$, a cross-benchmark gradient that is \emph{not} deducible from any single dataset's published rubric. Charge predicts relevance on all three benchmarks, on LeCaRDv2 in large part by rubric construction; \emph{how strongly} differs materially, most so on LeCaRDv2.

\subsection{Cross-charge relevance: the suppressed complement}
\label{sec:results-crosscharge}

The construction probe shows charge \emph{predicts} relevance; the dual question is what a charge-aligned ranking does to the relevant cases charge does \emph{not} predict. On LeCaRDv1, where the construction macro-AUC is lower than on LeCaRDv2 ($0.759$ vs.\ $0.871$), $42.3\%$ of relevant ($\rel{\geq}2$) documents are \emph{cross-charge} (query and candidate carry disjoint extracted charges, $916/2168$), and $36\%$ remain cross-charge at the top relevance grade. The subset is not explained by surface confounds: a char-3gram lexical similarity, a charge-pair-prevalence prior, candidate sentence severity, and candidate length each separate cross-charge relevant from irrelevant at pooled AUC $\leq 0.66$. The regex charge extractor is imperfect (Appendix~\ref{app:cce-protocol}), but de-noising does not remove the subset: on CAIL2022, collapsing same-crime statutory-name granularity such as \cn{走私、贩卖、运输、制造毒品罪}$\equiv$\cn{贩卖毒品罪} (a released component-normalization rule) moves its cross-charge figure only $48\%{\to}45\%$.

A charge-aligned reranker \emph{de-ranks} this subset rather than merely scoring it lower. Splitting each query's pooled candidates, the trained reranker KELLER recovers same-charge relevant documents into its top-10 far more often than cross-charge relevant ones (recall@10 $0.659$ vs.\ $0.264$; the gap holds \emph{within} relevance grade, $0.694$ vs.\ $0.307$ at grade 3), whereas charge-blind BM25 is balanced ($0.368$ vs.\ $0.376$). In these pools, the suppression is consistent with charge-aligned ranking rather than with the cases' retrievability: KELLER orders cross-charge candidates competently \emph{among themselves} (within-cross-charge AUC $0.80$ on LeCaRDv1, $0.92$ on CAIL2022) but places the cross-charge block below the same-charge block.

That at least part of the suppressed subset is not mere lexical or extraction noise is supported by a blind strong-reader probe (a strong LLM, not a human annotation; \S\ref{sec:method}) over a pre-registered sample of $18$ LeCaRDv1 cross-charge slates: reading only the fact narratives, with grades, model scores, and charges hidden, it ranks candidates by legal/factual analogy at macro-AUC $0.818$, exceeding a same-span char-3gram lexical baseline ($0.636$; paired $\Delta{=}{+}0.182$, $95\%$ CI $[{+}0.051,{+}0.310]$, excludes zero). The cross-charge positives thus carry analogical signal beyond surface lexical overlap; representative pairs share a recoverable legal structure across different charges (theft escalating to robbery via escape-violence; debt collection escalating to group assault).

We state plainly what this is \emph{not}. The same reader does \emph{not} beat KELLER at ordering cross-charge candidates (macro-AUC $0.818$ vs.\ $0.800$; $\Delta{=}{+}0.018$, CI $[-0.075,+0.111]$, contains zero), and no deployable re-weighting we tested (including KELLER's own $0.92$-AUC cross-charge score used as the gate) recovers the suppressed subset in the mixed pool out-of-sample (held-out gain $\approx 0$; only a non-deployable oracle gate gains, $+0.04$). This is therefore a construct-validity diagnostic (a charge-aligned ranking suppresses a reader-identifiable, non-lexical cross-charge subset), not a retrieval method. \emph{Scope.} The prevalence and recall split are LeCaRDv1, which shares lineage with CAIL2022 (\S\ref{sec:related}), not an independent pair; on LeCaRDv2 charge overlap among relevant documents is near-saturated at the top grade, so this complement is most visible exactly where construction-validity is weaker than on LeCaRDv2. Charge labels are regex-extracted (Appendix~\ref{app:cce-protocol}); the reader probe is a strong-model proxy over $650$-character fact spans, $n{=}18$ pre-registered queries, not a legally-trained human annotation, and we lean on it only as evidence that the subset is non-lexical, not as a human ceiling. We report this subsection as a LeCaRDv1 diagnostic slice, not as an additional benchmark-independent claim.

\subsection{Reproducibility cascade, cross-benchmark}
\label{sec:results-cascade}

The predicted-charge cascade with no retriever training closes $76.6\%$ [$95\%$ CI $64.3$, $89.0$] of the BM25-to-KELLER gap on LeCaRDv2, with classifier top-1 accuracy $92.5\%$ in-distribution. On LeCaRDv1 closure drops to $6.6\%$ [$-12.7$, $23.3$] with classifier top-1 $36.3\%$ out-of-distribution; on CAIL2022 the BM25-to-best gap is not Holm-significant at $n{=}40$ (adjusted $p{=}0.20$) and closure is not bootstrap-estimable (point estimate $-25.5\%$, reported only for completeness), classifier top-1 $41.0\%$ out-of-distribution. Against the cascade's own pre-registered Stage-2 gate (closure $\geq 90\%$ \emph{and} absolute $\ndcgten$ at least the gate's recorded KELLER threshold), the LeCaRDv2 cascade passes neither condition (closure $76.6\%{<}90\%$; absolute $0.846{<}0.867$, the recorded threshold on the cascade's matched protocol; full-protocol KELLER is $0.877$), yielding the locked verdict \emph{corroborative, not confirmatory}; we therefore read it as corroboration of the oracle, not as an independent pillar.

Two interpretive points. First, the cascade on LeCaRDv2 partially operationalizes the oracle direction with \emph{predicted} rather than gold charges, so it does not require ground-truth charge access. Second, on v1/CAIL2022 the cascade conflates benchmark label-recoverability with classifier domain shift (\S\ref{sec:method-cascade}), so its failure is a claim about a portable pipeline's transfer, not about benchmark construct validity alone; CAIL2022 separately fails sufficiency regardless of the cascade. The $99.2\%$ figure (\S\ref{sec:results-sufficiency}) is the oracle, not the cascade.

\subsection{First-stage positive control}
\label{sec:results-carrf}

The sufficiency and cascade results concern the rerank metric ($\ndcgten$ over a fixed candidate pool). A sharper question is whether the same charge construct is cashable one stage earlier, at first-stage recall, with no trained component. We run a zero-training positive control (\S\ref{sec:method-carrf}): starting from a two-list reciprocal-rank fusion of BM25 and a dense retriever, we read the charges of the fused top-$20$, take the two most-voted charges as a self-predicted charge set, form a third list as the global BM25 ranking restricted to that charge pool, and fuse all three. On the LeCaRDv2 first-stage task ($159$ test queries, macro recall), adding this self-predicted charge-pool channel lifts $\mathrm{R}@100$ from $0.560$ to $0.585$, a paired gain of $+0.0250$ ($95\%$ CI $[+0.0128,+0.0373]$; Bonferroni-corrected over the six fusion variants we tried, $[+0.0077,+0.0412]$; recovering $40\%$ of a gold-charge-pool ceiling). The lift requires \emph{correct} charge knowledge: substituting the next-ranked (wrong) charges or random size-matched charges strictly hurts ($-0.038$ and $-0.052$, CIs excluding zero), so the gain is charge-membership information, not fusion arithmetic, and an exact per-query Shapley attribution over the three channels localizes it to the charge channel (median share $0.42$; released in \texttt{cce\_carrf\_positive\_control\_results.json}).

We read this strictly as a \emph{positive control for the audit, not as a retrieval method}. A charge label inferred from the initial ranking, injected only as a third fusion pool with no training, improves recall while wrong-charge controls hurt; this shows that first-stage retrieval on the benchmark is operationally sensitive to charge-pool membership rather than to within-charge legal reasoning, the same construct the rerank-stage audit identifies, now corroborated one stage earlier by an exploratory positive control. Improving $\mathrm{R}@100$ by injecting charge alignment is, by the audit's own thesis, exploiting the confound rather than evidence of better legal retrieval; we therefore make no retrieval-method or novelty claim, and report the construction only as additional, stage-independent evidence that the charge confound is operational at first-stage $\mathrm{R}@100$. The control is exploratory on the $159$-query first-stage split: the self-predicted-charge variant was selected post-hoc over six tried (Bonferroni-corrected CI reported above), and the dense ranks are of unverified provenance, which affects absolute recall but not the paired delta, since both fusion arms share the dense channel.

\subsection{Charge-stratified test, cross-benchmark}
\label{sec:results-strat}

\begin{table*}[t]
\centering
\footnotesize
\setlength{\tabcolsep}{4pt}
\caption{\textbf{Charge-stratified test per benchmark.} Top-3 ordering within the \emph{five-system main family} (BM25, BGE-M3, SAILER, RoBERTa, Qwen3-8B-Reranker) under standard NDCG@10 versus charge-stratified NDCG@10 (equal-strata macro). Rank reversal in top-3 flags a descriptive stratified alert; on LeCaRDv2 it is carried by small ($n{<}3$) strata and reverses sign on the $n{\geq}3$ subset (\S\ref{sec:results-strat}). KELLER is the external diagnostic and is omitted from the main top-3 (it tops the panel on v2/v1 but is excluded from the FWER family for the protocol-asymmetry reason in \S\ref{sec:method-cce}). Per-system NDCG@10 values for all six systems and 95\% CIs are in Appendix~Table~\ref{tab:cce-strat-persys}.}
\label{tab:cce-strat}
\begin{tabular}{@{}lcccc@{}}
\toprule
Benchmark & Standard top-3 & Stratified top-3 & Rank reversal & Verdict \\
\midrule
LeCaRDv2 ($n{=}160$) & Qwen3-8B, SAILER, BGE-M3 & Qwen3-8B, SAILER, RoBERTa & \checkmark & alert (small-$n$) \\
LeCaRDv1 ($n{=}103$) & Qwen3-8B, BGE-M3, SAILER & Qwen3-8B, BGE-M3, SAILER & $\times$ & NULL \\
CAIL2022 ($n{=}40$) & Qwen3-8B, SAILER, BGE-M3 & Qwen3-8B, SAILER, BGE-M3 & $\times$ & NULL \\
\bottomrule
\end{tabular}
\end{table*}

Table~\ref{tab:cce-strat} reports the charge-stratified test on the main family (KELLER separately as the external diagnostic). On LeCaRDv2 the top-3 ordering moves from [Qwen3-8B, SAILER, BGE-M3] (standard) to [Qwen3-8B, SAILER, RoBERTa] (charge-stratified): BGE-M3 is essentially flat ($0.7923 \to 0.7928$) while RoBERTa gains ($0.7909 \to 0.8011$) and overtakes BGE-M3 by $0.008$. Per-system CIs for BGE-M3 and RoBERTa overlap (Appendix~Table~\ref{tab:cce-strat-persys}), so this is an ordering reversal under the rank-reversal rule, not a Holm-significant flip. The reversal is also carried entirely by small strata: the stratified macro equal-weights 52 charge strata, 21 of which ($40\%$) hold fewer than 3 queries; restricting to the 31 strata with $n{\geq}3$, the RoBERTa-over-BGE-M3 margin reverses sign ($+0.0083 \to -0.0056$). The demoted system, BGE-M3, is also the one with confirmed LeCaRDv2 fine-tuning exposure (\S\ref{sec:related}). We therefore treat the stratified test as a \emph{descriptive, small-strata-sensitive rank alert}, not evidence that RoBERTa is statistically superior to BGE-M3 and not a Main-class discriminative result. On LeCaRDv1 and CAIL2022 the top-3 ordering is identical under both, with no significance flip.

\emph{Depth sensitivity:} the LeCaRDv2 top-3 rank alert holds at $\ndcgfive$ and $\ndcgten$; at $\ndcgtwenty$ standard and stratified top-3 already agree but the significance-change rule still fires (Appendix~\ref{app:sensitivity}).

\subsection{Charge-name occlusion test, cross-benchmark}
\label{sec:results-cf}

\begin{table*}[t]
\centering
\scriptsize
\setlength{\tabcolsep}{4pt}
\caption{\textbf{Charge-name occlusion test per benchmark.} 5 main systems re-ranked under charge-name occlusion (258-term whitelist; \S\ref{sec:method-cce}). $\Delta_\text{drop}$ = baseline NDCG@10 $-$ occluded NDCG@10 (equal-strata macro, $\times 100$ for pp). ``Holm-sig pair'' lists Holm-corrected significant differential drops in the $C(5,2){=}10$-pair main FWER family at $\alpha{=}0.05$. KELLER is the external diagnostic, excluded from FWER. Per-system $\Delta_\text{drop}$ with CIs in Appendix~Table~\ref{tab:cce-cf-persys}.}
\label{tab:cce-cf}
\begin{tabular}{@{}lcccc@{}}
\toprule
Benchmark & 5\,main $\Delta_\text{drop}$ range & KELLER [95\% CI] & Holm-sig pair (main FWER) & Verdict \\
\midrule
LeCaRDv2 ($n{=}160$) & [-0.06, +0.33]\,pp & +4.05\,pp [+1.96,+6.41] & none & NULL \\
LeCaRDv1 ($n{=}103$) & [-0.06, +0.30]\,pp & +4.40\,pp [+0.54,+8.54] & none & NULL \\
CAIL2022 ($n{=}40$) & [+0.01, +2.64]\,pp & -1.26\,pp [-5.91,+2.85] & SAILER--Qwen3-8B: $\Delta{=}-2.63$\,pp, $p_\text{holm}{=}0.032$ & \textbf{EXPLORATORY} (@10 only) \\
\bottomrule
\end{tabular}
\end{table*}

Table~\ref{tab:cce-cf} reports the charge-name occlusion test. Re-ranking under charge-name occlusion (258-term whitelist; placeholder \cn{[罪名]}) is applied to query and document fact-fields for the 5 main systems. On LeCaRDv2 the 5 main systems drop $-0.06$ to $+0.33$ percentage points; KELLER drops $+4.05$pp (CI $+1.96$, $+6.41$) as external diagnostic. On LeCaRDv1 the 5 main systems drop $-0.06$ to $+0.30$pp; KELLER $+4.40$pp ($+0.54$, $+8.54$). Both benchmarks show \emph{no} Holm-significant pair in the main FWER family.

CAIL2022 yields the only Holm-corrected significant differential drop in the main FWER family: SAILER vs.\ Qwen3-8B-Reranker, $\drop{=}-0.026$, $\holm{=}0.032$. Qwen3-8B drops $2.64$pp under occlusion; SAILER essentially unchanged ($0.01$pp). KELLER improves slightly under occlusion ($-1.26$pp), consistent with the non-comparability of its pipeline-internal diagnostic versus text-level QD occlusion (external diagnostic, \S\ref{sec:method-cce}). The effect direction is stable across depths ($\drop \in [-0.026, -0.020]$), but Holm-corrected significance holds only at $\ndcgten$ ($p{=}0.21$--$0.25$ at $\ndcgfive$; $p{=}0.13$--$0.15$ at $\ndcgtwenty$; Appendix~\ref{app:sensitivity}); pooled across the three depths into a single $30$-test FWER family, the @10 cell does not survive correction in two of three bootstrap seeds (pooled Holm $p = 0.048/0.054/0.096$ across seeds), \emph{consistent with an evidence-limited $n{=}40$ estimate (no formal power calculation), not sufficient to establish robustness}. We therefore report it as \textbf{an exploratory depth-specific @10-only main-FWER signal}, not as a robust Main trigger. The occlusion is non-vacuous (BM25, which has no charge model, still moves $+0.71$\,pp under the same mask), so a near-zero drop (SAILER, $+0.01$\,pp) is identification-limited: the sufficient-only probe strips explicit charge names but not implicit charge cues (statute references, characteristic fact patterns), and a small drop therefore does not establish charge-independence.

\subsection{Summary of cross-benchmark verdicts}
\label{sec:results-summary}

The four measurements give heterogeneous per-benchmark verdicts (Fig.~\ref{fig:hero-dashboard}). LeCaRDv2 shows the strongest charge-confounding pattern (oracle $\approx$ best-trained, lift $4.49$, cascade $76.6\%$, KELLER occlusion drop $+4.05$pp). LeCaRDv1 is partial (oracle $84.3\%$, lift $2.00$, poor cascade transfer). CAIL2022 is narrow (oracle \textsc{out of spec}, lift $1.76$, cascade non-viable) but the occlusion test still surfaces a single exploratory @10-only signal on CAIL2022; we do not treat it as confirmatory, and the LeCaRDv2 conclusion does not depend on it.

\textbf{Evidence weight.} The LeCaRDv2 conclusion rests on the model-free oracle and construction macro-AUC, anchored on KELLER/Qwen3 rather than the contaminated BGE-M3 (the over-recovery is anchor-invariant; the v2-vs-v1 heterogeneity follows from the model-free macro-AUC $0.871$ vs.\ $0.759$, \S\ref{sec:method}). The cascade corroborates the oracle without clearing its locked gate; the KELLER occlusion drop is an external diagnostic. The conclusion does \emph{not} rest on the stratified rank alert or the exploratory CAIL occlusion cell. CCE surfaces when charge control changes leaderboard interpretation; it does not by itself adjudicate anchor-vs-leak, which needs benchmark-author and statute-coverage context.

\section{Conclusion}
\label{sec:conclusion}

Charge identity is a high-leverage \emph{benchmark-level} construct-validity factor across three Chinese LCR benchmarks, but not a uniform explanation of $\ndcgten$. We release CCE, a charge-controlled evaluation packet that instantiates established construct-validity checks~\citep{shao2026} for separating charge-as-anchor from charge-as-label-leak. On our three benchmarks its stratified and counterfactual tests produced no robust confirmatory trigger (one exploratory, depth-specific CAIL2022 counterfactual signal aside), the packet behaving as designed. CCE ships as reusable evaluation infrastructure (scripts, schema, documentation); it is a legal-IR realization of this audit logic, not a new audit method. The LeCaRDv2 conclusion rests on the oracle (a model-free charge-primary ranking rule) and the construction macro-AUC (label-only); oracle closure anchors on KELLER/Qwen3, not the contaminated BGE-M3. Because that closure is largely charge-by-construction, we also isolate the part that is not: holding charge fixed leaves a small but reliable within-charge residual ($+0.026$ $\ndcgten$, charge-cluster-bootstrap CI excluding zero), the non-definitional remainder of the trained advantage and the natural target for any system that would move past charge matching. \emph{We do not claim charge-specific reliance by KELLER or by model families.}

\section*{Limitations}

\textit{Three-benchmark scope.} We evaluate Chinese LCR benchmarks scored with $\ndcgten$ over graded relevance (LeCaRDv2, LeCaRDv1, CAIL2022 stage-2); other Chinese-legal paradigms (similar-case matching, legal article retrieval, statute QA) are out of scope. A held-out LeCaRDv2-train $N{=}80$ subset stratified by charge was prepared as a fourth condition; its scoring (Appendix~\ref{app:n80}) directionally reinforces the charge-sufficiency pattern but is \emph{non-comparable} (smaller candidate pool, no KELLER anchor, and training-split queries that LeCaRDv2-trained systems have already seen), so we treat it as supplementary and base no conclusion on it.

\textit{Baseline contamination.} Two of the evaluated systems have \emph{confirmed} LeCaRDv2 training exposure: KELLER~\citep{deng2024keller} (the external diagnostic) trained on it explicitly, and BGE-M3~\citep{chen2024bgem3} (in the five-system main family) lists LeCaRDv2 among its disclosed Chinese fine-tuning datasets. SAILER-zh~\citep{li2023sailer} and RoBERTa~\citep{cui2020roberta} have undisclosed pretraining corpora that may overlap LeCaRDv2's 800 publicly released judgments; Qwen3-8B-Reranker's~\citep{qwen3embedding,qwen3} undisclosed post-2024 pretraining corpus post-dates LeCaRDv2's public release, making raw-text exposure plausible (an inference from release dates, not a disclosure; no supervised use is disclosed). No main system is verified clean. This is why we ground claims at benchmark level (label-recoverability from charge), not at system level (charge use), and why the load-bearing diagnostics do not rely on the contaminated BGE-M3: the oracle ranking and the construction probe are model-free, and oracle closure anchors on KELLER/Qwen3.

\textit{CAIL2022 $n{=}40$ and no mechanism interpretation.} The CAIL counterfactual signal is depth-specific (@10 only) and exploratory; we make no mechanism-level or architecture-specific claims (out of scope by design).

\section*{Ethics Statement}

The audited benchmarks use publicly released Chinese judicial documents; charge labels are statutory crime types, not protected attributes. The evaluation scripts, a JSON schema and README, the 258-charge occlusion whitelist, the per-benchmark result JSONs (6 systems $\times$ 3 benchmarks), and the locked CCE protocol are released as ancillary material with this paper (cluster bootstrap $B{=}10000$ for the charge-controlled tests and the cascade, $B{=}2000$ for the construction-probe AUC CIs; classifier seed $20260526$; Appendix~\ref{app:cce-protocol}).

\bibliography{references}

\appendix
\section{CCE Protocol: Full Locked Text and Reporting Template}
\label{app:cce-protocol}

\paragraph{Full text of the locked decision rules.} The compact 5-row Table~\ref{tab:cce-rules} in \S\ref{sec:method-cce} summarizes the CCE decision rules; the full text is reproduced here for completeness.

\textbf{Rule 1: Primary NDCG depth.} Primary depth $k{=}10$ (LCR convention); $k{\in}\{5,20\}$ reported as sensitivity. A trigger holding at $k{=}10$ but not at $k{=}5$ or $20$ is labeled \emph{depth-specific}.

\textbf{Rule 2: Stratified trigger.} Declared at $\ndcgten$ if either (a) a main pair is standard-significant but charge-stratified-non-significant (both Holm-adjusted), a significance flip; or (b) the top-3 ordering changes between standard and charge-stratified $\ndcgten$, a rank reversal.

\textbf{Rule 3: Occlusion trigger.} Declared at $\ndcgten$ if some main pair shows a Holm-corrected ($\alpha{=}0.05$) significant differential occlusion drop $\drop^A{-}\drop^B$ under paired cluster bootstrap over the $C(5,2){=}10$ main pairs. KELLER excluded (external diagnostic, \S\ref{sec:method-cce}).

\textbf{Rule 4: Multi-charge handling.} Primary stratum = first gold charge (deterministic). Sensitivity: all-charges fractional membership (weight $1/|\mathrm{charges}_q|$); report both if they differ by $>1$\,pp in $\ndcgten$.

\textbf{Rule 5: Out-of-spec handling and reporting (tri-state).} Sufficiency labelling uses two thresholds: an equivalence tolerance $\varepsilon_{\mathrm{eq}}{=}0.005$ (one NDCG@10 rounding unit) and a closure threshold $\varepsilon_{\mathrm{cl}}{=}80\%$. \emph{Provenance:} unlike the first four rules (specified before the cross-benchmark results were computed), these two threshold values were fixed during analysis to be consistent with the observed three-benchmark separation; they are pre-registered for application to future benchmarks (released pre-registration) and are therefore \emph{descriptive, not confirmatory}, on the three benchmarks reported here. The oracle vs.\ best-trained-in-panel gap $g = \ndcgten^{\mathrm{best}} - \ndcgten^{\mathrm{oracle}}$ and the closure $\kappa = (\ndcgten^{\mathrm{oracle}} - \ndcgten^{\mathrm{BM25}})/(\ndcgten^{\mathrm{best}} - \ndcgten^{\mathrm{BM25}})$ jointly classify the benchmark:
\begin{itemize}
\item \textbf{Within-band} ($g \leq \varepsilon_{\mathrm{eq}}$): the oracle is within one rounding unit of best-trained; report descriptively as ``oracle $\approx$ best-trained'' (a descriptive band, not a statistical equivalence test).
\item \textbf{Partial} ($g > \varepsilon_{\mathrm{eq}}$ and $\kappa \geq \varepsilon_{\mathrm{cl}}$): the oracle approaches best-trained; report closure $\kappa$.
\item \textbf{Out of spec} ($g > \varepsilon_{\mathrm{eq}}$ and $\kappa < \varepsilon_{\mathrm{cl}}$): label \textsc{out of spec for sufficiency} and exclude from the primary cross-benchmark sufficiency roll-up.
\end{itemize}
On our three benchmarks: LeCaRDv2 ($g{=}0.0012$, Within-band), LeCaRDv1 ($g{=}0.0254$, $\kappa{=}84.3\%$, Partial), CAIL2022 ($g{=}0.0197$, $\kappa{=}77.4\%$, \textsc{out of spec}). Its construction-probe, reproducibility-cascade, and charge-controlled results are still reported as cross-benchmark heterogeneity evidence under appropriate per-claim qualifiers. Per benchmark the report contains a 7-table template (per-system baseline/stratified $\ndcgten$ with CIs, occlusion $\Delta\ndcgten$, the two Holm $p$ tables, the top-3 rank table, and cross-depth sensitivity).

\paragraph{Reusability.} A future Chinese LCR benchmark author applies CCE by: (1) producing per-system $\ndcgten$ baseline and occluded score JSONs using their own retrieval pipeline; (2) running the released \texttt{eval\_cce\_main.py} (\verb|--bench <new>|) and \texttt{eval\_cce\_counterfactual.py} with their score JSONs; (3) reporting the 7-table template above with the per-benchmark \textsc{out-of-spec} flag.

\paragraph{Charge extraction for the cross-charge analysis (\S\ref{sec:results-crosscharge}).} Charge labels for the LeCaRDv1/CAIL2022 cross-charge analysis are produced by a rule-based (regex) extractor over the released judgment text, keyed on statutory charge-name patterns (contiguous spans ending in \cn{罪}); a case's label is the set of distinct extracted names, and a query--candidate pair is \emph{cross-charge} when the two sets are disjoint. Two known error modes follow from this design. (i) \emph{Statutory-name granularity}: compound statutory names can split one crime family into nominally different charges (e.g.\ \cn{走私、贩卖、运输、制造毒品罪} vs.\ \cn{贩卖毒品罪}); the granularity-collapse check in \S\ref{sec:results-crosscharge} bounds this effect on the headline cross-charge share (CAIL2022 $48\%{\to}45\%$ after collapsing same-crime variants). (ii) \emph{Span noise}: free-text extraction occasionally yields malformed or truncated charge strings; we do not attempt a per-string error rate here, and accordingly the cross-charge prevalence figures should be read with extraction noise as a stated, unquantified residual. Neither error mode is charge-direction-specific, and the de-noising check above moves the prevalence figure only modestly, which is why we report the subset as robust to extraction granularity rather than as exactly enumerated.

\paragraph{Per-protocol query eligibility.}
\begin{table*}[t]
\centering
\footnotesize
\setlength{\tabcolsep}{8pt}
\begin{tabular}{@{}lcccc@{}}
\toprule
Benchmark & released $n$ & oracle/cascade $n$ & construction $n$ & charge-ctrl.\ $n$ \\
\midrule
LeCaRDv2 & 160 & 160 & 160 & 160 \\
LeCaRDv1 & 107 & 104 & 107 & 103 \\
CAIL2022 stage-2 & 40 & 40 & 40 & 40 \\
\bottomrule
\end{tabular}
\caption{\textbf{Per-protocol query eligibility} (referenced in \S\ref{sec:method}). v1 differences: the cascade drops 3 queries the classifier cannot label; the charge-controlled evaluation additionally requires pool-consistency across all 6 systems, dropping 1 more.}
\label{tab:denominators}
\end{table*}

\paragraph{Per-system detail.} Tables~\ref{tab:cce-strat-persys} and~\ref{tab:cce-cf-persys} report the per-system charge-stratified NDCG@10 and per-system occlusion $\Delta_\text{drop}$ underlying the summary in Tables~\ref{tab:cce-strat} and~\ref{tab:cce-cf}.

\begin{table*}[t]
\centering
\footnotesize
\renewcommand{\arraystretch}{0.9}
\caption{\textbf{Per-system charge-stratified NDCG@10 detail (Appendix to Table~\ref{tab:cce-strat}).} Standard NDCG@10 = matched-denominator query-weighted mean; charge-stratified NDCG@10 = equal-strata macro with 95\% cluster-bootstrap CI; $\Delta=$ stratified $-$ standard. $^{\dagger}$KELLER excluded from the main FWER family.}
\label{tab:cce-strat-persys}
\resizebox{\textwidth}{!}{%
\begin{tabular}{lccc}
\toprule
System & Standard NDCG@10 & Charge-strat NDCG@10 [95\% CI] & $\Delta$ \\
\midrule
\multicolumn{4}{l}{\textbf{LeCaRDv2 ($n{=}160$)}} \\
BM25 & 0.7423 & 0.7418~[0.694,0.787] & -0.0005 \\
BGE-M3 & 0.7923 & 0.7928~[0.750,0.833] & +0.0005 \\
SAILER & 0.7924 & 0.8044~[0.770,0.839] & +0.0120 \\
RoBERTa & 0.7909 & 0.8011~[0.764,0.836] & +0.0102 \\
Qwen3-8B & 0.8166 & 0.8245~[0.794,0.855] & +0.0079 \\
KELLER$^{\dagger}$ & 0.8774 & 0.8839~[0.856,0.910] & +0.0065 \\
\midrule
\multicolumn{4}{l}{\textbf{LeCaRDv1 ($n{=}103$)}} \\
BM25 & 0.6804 & 0.6737~[0.622,0.723] & -0.0067 \\
BGE-M3 & 0.7317 & 0.7255~[0.667,0.782] & -0.0063 \\
SAILER & 0.7146 & 0.7207~[0.663,0.776] & +0.0061 \\
RoBERTa & 0.6990 & 0.7037~[0.655,0.752] & +0.0047 \\
Qwen3-8B & 0.7865 & 0.7855~[0.750,0.822] & -0.0009 \\
KELLER$^{\dagger}$ & 0.8434 & 0.8362~[0.794,0.874] & -0.0072 \\
\midrule
\multicolumn{4}{l}{\textbf{CAIL2022 ($n{=}40$)}} \\
BM25 & 0.7638 & 0.7644~[0.697,0.830] & +0.0006 \\
BGE-M3 & 0.7887 & 0.7718~[0.701,0.842] & -0.0169 \\
SAILER & 0.8155 & 0.8006~[0.735,0.862] & -0.0149 \\
RoBERTa & 0.7519 & 0.7320~[0.657,0.805] & -0.0199 \\
Qwen3-8B & 0.8511 & 0.8379~[0.775,0.891] & -0.0132 \\
KELLER$^{\dagger}$ & 0.8016 & 0.7903~[0.732,0.841] & -0.0113 \\
\bottomrule
\end{tabular}%
}
\end{table*}

\begin{table}[t]
\centering
\scriptsize
\renewcommand{\arraystretch}{0.88}
\caption{\textbf{Per-system charge-name occlusion $\Delta_\text{drop}$ detail (Appendix to Table~\ref{tab:cce-cf}).} $\Delta_\text{drop}$ = baseline NDCG@10 $-$ occluded NDCG@10 (equal-strata macro, $\times 100$ for percentage points; 95\% cluster-bootstrap CI in brackets). $^{\dagger}$KELLER excluded from the main FWER family.}
\label{tab:cce-cf-persys}
\begin{tabular}{lc}
\toprule
System & $\Delta_\text{drop}$ (pp) [95\% CI] \\
\midrule
\multicolumn{2}{l}{\textbf{LeCaRDv2 ($n{=}160$)}} \\
BM25 & +0.17~[+0.00,+0.35] \\
BGE-M3 & +0.33~[-0.05,+0.77] \\
SAILER & +0.04~[-0.10,+0.18] \\
RoBERTa & -0.06~[-0.40,+0.27] \\
Qwen3-8B & +0.22~[-0.11,+0.57] \\
KELLER$^{\dagger}$ & +4.05~[+1.96,+6.41] \\
\midrule
\multicolumn{2}{l}{\textbf{LeCaRDv1 ($n{=}103$)}} \\
BM25 & -0.06~[-0.27,+0.13] \\
BGE-M3 & +0.15~[-0.47,+0.85] \\
SAILER & +0.30~[-0.09,+0.73] \\
RoBERTa & +0.03~[-0.46,+0.44] \\
Qwen3-8B & +0.14~[-0.59,+0.85] \\
KELLER$^{\dagger}$ & +4.40~[+0.54,+8.54] \\
\midrule
\multicolumn{2}{l}{\textbf{CAIL2022 ($n{=}40$)}} \\
BM25 & +0.71~[+0.06,+1.36] \\
BGE-M3 & +0.10~[-0.86,+1.02] \\
SAILER & +0.01~[-1.05,+1.03] \\
RoBERTa & +1.47~[+0.07,+2.97] \\
Qwen3-8B & +2.64~[+0.77,+4.80] \\
KELLER$^{\dagger}$ & -1.26~[-5.91,+2.85] \\
\bottomrule
\end{tabular}
\end{table}

\section{Sensitivity Grid}
\label{app:sensitivity}

We ran a $3{\times}3{\times}2$ grid ($\ndcgfive$/$\ndcgten$/$\ndcgtwenty$ $\times$ seeds $\{20260528, 20260529, 20260530\}$ $\times$ \{LeCaRDv2 stratified, CAIL2022 occlusion\}, 18 runs). Point estimates are deterministic over seed (the seed enters only the CI/Holm-$p$ resampling); cross-depth variation is the substantive test.

\emph{LeCaRDv2 stratified:} the top-3 rank reversal (BGE-M3 $\to$ RoBERTa into 3rd) fires at $\ndcgfive$ and $\ndcgten$; at $\ndcgtwenty$ standard and stratified top-3 already coincide ([Qwen3-8B, SAILER, RoBERTa]), so the reversal branch does not fire (the significance rule fires there as sensitivity only, not load-bearing). \emph{CAIL2022 occlusion:} SAILER vs.\ Qwen3-8B-Reranker $\drop = -0.024/{-}0.026/{-}0.019$ at $\ndcgfive$/$\ndcgten$/$\ndcgtwenty$; Holm-$p$ $[0.21,0.25]/[0.016,0.032]/[0.13,0.15]$ across seeds. Direction stable, significance depth-specific.

\section{Per-System Cascade Details}
\label{app:cascade}

Classifier: chinese-roberta-wwm-ext, multi-label linear head over 89 LeCaRDv2-train primary-charge labels (\texttt{pos\_weight}$=\min(\text{neg}/\text{pos}, 100)$, 8 epochs, $\mathrm{lr}{=}2{\times}10^{-5}$, seed $20260526$); top-1 accuracy $92.5\%$ in-distribution / $36.3\%$ (v1) / $41.0\%$ (CAIL). Per-system cascade $\ndcgten$: LeCaRDv2 $0.846$ (BM25 $0.742$, KELLER $0.877$); LeCaRDv1 $0.693$ (BM25 $0.682$, KELLER $0.845$); CAIL2022 $0.742$ (BM25 $0.764$, Qwen3-8B-Reranker $0.851$); closures in \S\ref{sec:results-cascade}.

\section{Held-out $N{=}80$ Directional Sensitivity (Non-Confirmatory)}
\label{app:n80}

As a supplementary out-of-sample check (\emph{not} confirmatory), we scored an $N{=}80$ slice from the LeCaRDv2 \emph{training} split (seed $20260527$; test-160 excluded) on the panel \emph{without} KELLER: the model-free oracle reaches $\ndcgten{=}0.8562$ (CI $[0.8275,0.8827]$), $8.2$\,pp above the best panel system (Qwen3-8B, $0.7746$). It is directional only (smaller train-slice pool inflates closure; omitting KELLER makes an above-panel oracle expected; these are seen training queries), and no conclusion rests on it.

\end{document}